\documentclass{article}

\usepackage{graphicx}
\usepackage{amssymb,amsfonts,amsmath}

\begin{document}

\title{Formation of buckminsterfullerene (C$_{60}$) in interstellar space}

\author{Olivier Bern\'e$^{1*}$ \and Alexander G. G. M. Tielens$^{1}$\\
 \\ $^{1}$Leiden Observatory, Leiden University, P.O. Box 9513, 
 \\NL- 2300 RA Leiden, The Netherlands
 \\$^*$Present address : IRAP, Universit\'e de Toulouse, CNRS
 \\ 9 Av. colonel Roche, BP 44346, F-31028 
 \\Toulouse, France
 \\ (olivier.berne@gmail.com)}

\maketitle

\begin{abstract}

\emph{Buckminsterfullerene} (C$_{60}$) was recently confirmed to be the largest molecule identified in space. However, it remains unclear how, and where this molecule is formed. It is generally believed that C$_{60}$ is formed from the build up of small carbonaceous compounds, in the hot and dense envelopes of evolved stars. Analyzing infrared observations, obtained by \emph{Spitzer} and \emph{Herschel}, we found that C$_{60}$ is efficiently formed in the tenuous and cold environment of an interstellar cloud illuminated by strong ultraviolet (UV) radiation fields. This implies that another formation pathway, efficient at low densities, must exist. Based on recent laboratory and theoretical studies, we argue that Polycyclic Aromatic Hydrocarbons are converted into graphene, and subsequently C$_{60}$, under UV irradiation from massive stars. This shows that alternative -- top-down -- routes are key to understanding the organic inventory in space.

\end{abstract}

%% When adding keywords, separate each term with a straight line: |
%\keywords{polycyclic aromatic hydrocarbons | graphene | fullerene | interstellar medium}

\section{Introduction}

The mid-infrared spectra of a variety of astrophysical objects are dominated by band emission (strongest at 3.3, 6.2, 7.7, 8.6 and 11.2 $\mu$m)
attributed to carbonaceous macromolecules, i.e. Polycyclic Aromatic Hydrocarbons~\cite{tie08} (PAHs). These molecules are large (30-100 C atoms), 
abundant ($\sim$ 5\% of the elemental carbon), and their ionization plays a key role in the energy balance of gas in the interstellar 
medium and in protoplanetary disk. In addition to PAH bands, infrared signatures observed at 7.0, 8.5, 17.4 and 19.0 $\mu$m have been reported recently~\cite{cam10,sel10}, 
and found to coincide precisely with the emission of  buckminsterfullerene (C$_{60}$)~\cite{kro85}, a cage-like carbon molecule. This detection
heralds the presence of a rich organic inventory and chemistry in space. However, observed abundances of C$_{60}$ challenge the standard 
ion-molecule or grain-surface chemistry formation routes, which build up molecules from small to large in the interstellar medium. For that reason, 
it has been suggested that  C$_{60}$ is formed in the hot and dense envelopes of evolved stars~\cite{goe92, che00, pas00} in processes alike to 
those found in sooty environments~\cite{kro88,hea92,hun98,irl06}, and eventually, is ejected in space. Yet, this scenario faces the
problem that it has a limited efficiency~\cite{che00}. PAHs and C$_{60}$ are known to co-exist in the interstellar medium~\cite{sel10}, 
however, so far, the connection between PAHs  and C$_{60}$ --and in particular the possibility to go from one compound  to the other in space-- 
has not been investigated. In this paper, we present a study of PAH and C$_{60}$ chemical evolution in the NGC 7023 nebula,
using \emph{Spitzer} \cite{wer04} and \emph{Herschel} \cite{pil10} infrared observations.

\subsection{Observational results}

\subsection{Infrared observations of the NGC 7023 nebula}
Earlier \emph{Spitzer}  observations of the NGC 7023 reflection nebula have revealed a chemical evolution of PAHs: deep 
in the cloud, emission is dominated by PAH clusters, which evaporate into free-flying PAHs when exposed to the UV radiation from the 
star~\cite{rap05,ber07,bre05}. There, gaseous PAHs are, in turn, ionized. While the neutral PAHs are dominated by zig-zag edges -- as 
demonstrated by the strong C-H solo out-of-plane modes --, the ions have an arm-chair molecular structure, characterized by strong 
duo out-of-plane modes~\cite{ros11}. In regions closest to the star, the presence of C$_{60}$ in the neutral state  
is evidenced by \emph{Spitzer} observations (Fig. 1). 
New \emph{Herschel} observations provide a measurement of dust emission in the same region, at high angular resolution (Fig. 1). 
This measurement can be used to derive the integrated intensity radiated by the nebula which can be used as a calibrator, to convert 
the \emph{Spitzer} observations of the PAHs and C$_{60}$ bands into absolute chemical abundances of these species, allowing a 
quantitive study of PAH and C$_{60}$ chemical evolution.

\subsection{Measurement of the far infrared integrated intensity of dust emission in the nebula}

The far infrared integrated intensity  $I_{FIR}$ was extracted by fitting the spectral energy distribution (SED) at each
position in the cross cut shown in Fig. 1. For these positions we have used the brightnesses as measured by \emph{Herschel}
PACS \cite{pog10} and SPIRE \cite{gri10} photometers. We have used the 70 and 160 $\mu$m channels of PACS and 
the 250 $\mu$m channel of SPIRE. This data is presented in details in \cite{abe10}. The modified blackbody function fit to 
these SEDs containing 3 spectral points is defined by:

\begin{equation}
I(\lambda, T)=K/\lambda^{\beta} \times B(\lambda, T),
\end{equation}

where $K$ is a scaling parameter, $\lambda$ is the spectral index, and $B(\lambda, T)$ is the planck function with
$\lambda$ the wavelength and $T$ the temperature. We have used a constant value of 1.8 for $\beta$ so that only
$T$ and $K$ are free parameters. The results of the fits to the observations are shown in Figure S1.
The temperatures we have derived from the fit of the data range between 
25 and 30 K. These values are in agreement with  those of \cite{abe10} for the same region. 
The peaking position of the modified blackbody function moves to shorter wavelengths (i.e. the grain temperature increases) 
when getting closer to the star, implying that we are indeed tracing matter inside the cavity and not material behind on the line of sight.
The far infrared integrated intensity, $I_{FIR}$, is then derived by integrating $I(\lambda, T)$ over frequencies. 

\subsection{Measurement of the integrated intensity of PAH and C$_{60}$ emission in the nebula}

The integrated intensity of PAH emission, $I_{PAH}$, is measured by fitting a PAH emission model
to the observed \emph{Spitzer} mid infrared low resolution spectrum. This data covers the 5 to 14 $\mu$m
range where most of the emission occurs. Because this spectral range contains the emission due to the vibration
of both C-C and C-H bonds, it is insensitive to ionization of PAHs. An example of this fitting
model is presented in \cite{ber08}. This tool provides the integrated intensity in the 
PAH bands as an output and takes care of continuum 
subtraction and extinction correction. The C$_{60}$ integrated intensity ($I_{C_{60}}$) extraction needs to be done separately 
for each band. We first measure the integrated intensity in the 19 and 17.4 band. 
The first step consists in extracting the intensity in the 19.0 
$\mu$m band ($I_{19.0}$) by fitting a gaussian and subtracting a local linear continuum.
The 17.4 $\mu$m band is contaminated by PAH emission, but this can be removed effectively.
The contribution of PAHs to the 17.4 band can be
estimated in this way: in the outer regions of the nebula, the 17.4 $\mu$m band is 100$\%$ due to 
PAHs (since no C$_{60}$ is detected there) and so is the 16.4 $\mu$m band. The 16.4 and 17.4 
$\mu$m band of PAHs are known to correlate. Indeed, in the outer regions of the nebula we find
$I_{17.4}^{PAH}\sim I_{16.4}\times 0.35$. Therefore, over the whole nebula, we can estimate the intensity
of the 17.4 $\mu$m band due to C$_{60}$ by $I_{17.4}^{C_{60}}\sim I_{17.4}-I_{16.4}\times 0.35$. Figure S2
shows the result of this process on the map. The 7.0 $\mu$m band is faint and usually hard to detect. 
In the regions closest to the star where both the 19 and 7.0 $\mu$m features are observed, we can 
calibrate the ratio of $I_{19.0}/I_{7.0}$. It is found to be relatively stable (at least in this small zone close to the star)  
and of the order of 0.4. Therefore we use $I_{7.0}=0.4 \times I_{19.0}$. The 8.5 $\mu$m band is undetectable 
because of the strong PAH band present at 8.6 $\mu$m, so we use the ratio provided in 
\cite{sel10} for 5 eV photons that is $I_{8.5}=0.4\times I_{19.0}$. The total C$_{60}$ IR emission
 is hence given by 
 $I_{C_{60}}=I_{19.0}+ I_{17.4}-I_{16.4}\times 0.35+I_{19.0}\times0.4+I_{19.0}\times0.4$.

\subsection{Derivation of PAH and C$_{60}$ abundances}

The method to derive the abundance of carbon locked in PAHs and C$_{60}$ from $I_{PAH}$, $I_{C_{60}}$ and $I_{FIR}$
is presented in details in \cite{tie05}. The fraction of carbon locked in PAHs and C$_{60}$ (respectively $f_{C}^{PAH}$ and $f_{C}^{C_{60}}$) 
per atom of interstellar hydrogen are given by:
\begin{equation}
f_{C}^{PAH}=0.23\times\left(\frac{7\times10^{-18}}{\sigma_{uv}^{PAH}}\right) \frac{R_{PAH}}{1-(R_{PAH})},
\end{equation}
and
\begin{equation}
f_{C}^{C_{60}}=0.23\times\left(\frac{7\times10^{-18}}{\sigma_{uv}^{C_{60}}}\right) \frac{R_{C_{60}}}{1-(R_{C_{60}})},
\end{equation}

where
\begin{equation}
R_{PAH}=\frac{I_{PAH}}{I_{PAH}+I_{C_{60}}+I_{FIR}},
\end{equation}
and
\begin{equation}
R_{C_{60}}=\frac{I_{C_{60}}}{I_{PAH}+I_{C_{60}}+I_{FIR}}.
\end{equation}

${\sigma_{uv}^{PAH}}$ and $\sigma_{uv}^{C_{60}}$ are the UV absorption cross sections of
PAHs and C$_{60}$. Following \cite{tie05}, we adopt ${\sigma_{uv}^{PAH}}=7\times10^{-18}$ cm$^{2}$
per C atom. There are no detailed measurements of the UV absorption cross section of C$_{60}$,
following \cite{sel10} we adopt the same value as for PAHs. 

\subsection{Evidence of C$_{60}$ formation in NGC 7023}

The results of the abundance variations within the nebula as derived using the above method are shown in 
Fig. 2. The C$_{60}$ abundance in the nebula is seen to increase from $1.4 \times 10 ^{-4}$\% to $1.5 \times 10^{-2}$\% of the elemental carbon 
abundance when approaching the star (Fig. 2). On the other hand, the abundance of Polycyclic Aromatic Hydrocarbons (PAHs) is seen 
to decrease, from 7.0\% to 1.8\% of the carbon (Fig. 2).  This shows that C$_{60}$ is being formed in the nebula while
PAHs are being destroyed or processed. The correlation of these variations with the increasing UV field strongly suggests
that UV photons control C$_{60}$ formation and PAH processing and destruction.

\section{The top-down model}

\subsection{Proposed scenario for the formation of C$_{60}$ in the ISM}

The formation of C$_{60}$ in the interstellar medium is unexpected and has several implications in our understanding of the formation 
process of this molecule. In the laboratory, the main formation route invoked for C$_{60}$ is the build-up from atomic carbon, small 
carbon clusters or rings~\cite{kro88,hea92,hun98,irl06}. In space, such processes are efficient in the hot (1500 K) and dense 
($n_H>10^{11}$ cm$^{-3}$) envelopes of evolved stars~\cite{che00}. In NGC 7023, the gas is $\sim$ 7 orders of magnitude less 
dense (see \cite{job10} and SI text), making this aggregation processes inefficient. Instead, we invoke photochemical processing of PAHs as an 
important route to form $C_{60}$. Upon UV irradiation, several channels for
fragmentation can be open depending on excitation energy; e.g., H-loss and C$_2$H$_2$ loss~\cite{eke97}.  However, experiments on 
small PAHs have shown that H-loss is by far the dominant channel~\cite{joc94} and that complete dehydrogenation i.e. 
graphene~\cite{nov05}  formation, is then the outcome of the UV photolysis process~\cite{job03,tie05, lep03, mon11}. 
In a second step, carbon loss followed by pentagon formation initiates the curling of the molecule~\cite{she10}. We envision that
this is followed by migration of the pentagons within the molecule, leading to the zipping-up of the open edges forming the 
closed fullerene~\cite{chu10} (Fig. 3 and SI Movie). Graphene formation through PAH photolysis can give rise to a rich chemistry. 
Besides the route towards fullerenes outlined above, fragmentation towards small cages, rings, and chains may also result.
The relative importance of isomerization and fragmentation will determine the carbon 
inventory delivered by the photo-chemical evolution of PAHs in space (Fig. 3). 

\subsection{PAH and Graphene stability in space}

Guided by experimental~\cite{job03} and 
theoretical~\cite{lep03} studies we have evaluated the stability of PAHs against dehydrogenation, and $C_{60}$
formation, in conditions appropriate for NGC 7023.

\subsubsection{Graphene formation}

Schematically,  the fragmentation process  can be written as,
\begin{equation}
{\rm PAH}\, +\, h\nu\, \leftrightarrow\, {\rm PAH^*}\, \rightarrow\, {\rm PAH_{-H}}\, +\, {\rm H}
\quad ,
\end{equation}
where PAH$^*$ is the excited species which can stabilize through emission of IR photons or through fragmentation and PAH$_{-H}$ is a dehydrogenated PAH radical. There are various ways to evaluate the unimolecular dissociation rate constant for this process (cf., \cite{tie05} sect 6.4). We will follow \cite{klo89} and write the rate constant in Arrhenius form,
\begin{equation}
k\left(E\right)\, =\, k_o\left(T_e\right)\, \exp\left[-E_o/kT_e\right]
\label{eqn:fragmentation rate}
\quad ,
\end{equation}
where $T_e$ is an effective excitation temperature, $E_o$, the Arrhenius energy describing the process, and the preexponential factor, $k_o$, depends on the interaction potential (in the reverse reaction). For PAHs, the internal excitation of the vibrational modes after the absorption of a UV photon can be well described by,
\begin{equation}
 T_m\, =\, 2000\, \left(\frac{E(\rm eV)}{N_c}\right)^{0.4}
\quad ,
\end{equation}
where $N_c$ is the number of C-atoms and $E$ is internal energy in eV (\cite{tie05} p. 184). Because typically the energy involved in these reactions is a fair fraction of the total energy in the system, a correction has to be made to this excitation temperature. The finite heat bath correction results in \cite{tie05, klo89},
\begin{equation}
 T_e\, =\, T_m\, \left(1\, -\, 0.2\, \frac{E_o}{E}\right)
\quad .
\end{equation}
The preexponential factor can be set equal to,
\begin{equation}
k_o\, =\, \frac{kT_e}{h}\, \exp\left[1\, +\, \frac{\Delta S}{R}\right]
\quad ,
\end{equation}
with $\Delta S$ the entropy change for which we will adopt 5 cal/K \cite{lep03}.  $k_o$ is then typically, $\simeq 3\times 10^{16}$ s$^{-1}$. The Arrhenius energy parameter, $E_o$, cannot be easily evaluated from theoretical calculations (\cite{bae96}). Here, we use a fit to the experimental fragmentation studies on small PAHs ($<24$ C-atoms; \cite{joc94}), which results in $E_o$ is 3.3 eV (\cite{tie05} p. 204). The probability for dissociation depends then on the competition between fragmentation and IR photon emission,
\begin{equation}
p_d\left(E\right)\, =\, \frac{k\left(E\right)}{k\left(E\right)\, +\, k_{ir}\left(E\right)}
\quad ,
\end{equation}
where $k_{ir}(E)$ is the IR emission rate at an internal energy $E$. For a highly excited PAH, $k_{ir}(E)$ is $\sim$ 1 s$^{-1}$. 
The total fragmentation rate is then,
\begin{equation}
k_{frag}\, =\, p_d\left(E\right)\, k_{uv}\left(E\right)
\quad ,
\end{equation}
where $k_{uv}\left(E\right)$ is the absorption rate of UV photons with energy, $E$.

The photochemically driven H-loss is balanced by reactions of atomic hydrogen with dehydrogenated PAHs; viz.,
\begin{equation}
{\rm PAH_{-H}}\, +\, H\, \rightarrow\, {\rm PAH} + \, h\nu
\quad .
\end{equation}
The rate of this reaction has been measured to be $k_a=1.5\times 10^{-10}$ cm$^3$ s$^{-1}$ for a number of small PAHs \cite{lep03,sno98}. We can define the dissociation parameter, $\psi\, =\, k_an_H/k_{frag}$. With $k_{uv}=7\times 10^{-10}\, N_c\, G_o$ ($G_o$ is the flux of UV photons in units of the interstellar Habing field \cite{hab68}, we have $\psi\simeq 0.2n_H/N_cG_op_d(E)$. The hydrogen coverage of interstellar PAHs is a very sensitive function of $\psi$ and a small increase of $\psi$ can change PAHs from fully hydrogenated to graphene \cite{tie05, sno98}. This is illustrated for the circumovalene, C$_{66}$H$_{20}$, in Figure S3. Thus, PAHs are fully hydrogenated if  $\psi$  is much less than $1$ and fully dehydrogenated if $\psi$ is much larger than $1$. Hence, 
\begin{equation}
\frac{G_o}{n_H}\, =\, \frac{0.2}{N_c}\,\left(1\, +\, \frac{k_{ir}(E)}{k(E)}\right)
\quad 
\end{equation}
provide a critical relation for the transformation of PAHs into graphene. We have evaluated this relation assuming an absorbed UV photon energy of 10 eV (Fig. 4).

\subsubsection{Fullerene formation} 

After formation of graphene, UV photoabsorption can lead to  loss of carbon from the skeleton. This fragmentation process competes  with stabilization through IR photon emission. The reaction rate is again given by eqn.~\ref{eqn:fragmentation rate}.  We adopt $\Delta S$ is 5 cal/K, resulting in $k_o\simeq3\times 10^{-16}$ s$^{-1}$, but the exact value is not critical. For $E_o$, we have adopted the Arrhenius energy (3.65 eV \cite{mic10}) derived from experiments on the C-loss from small catacondensed PAHs \cite{joc94}. The calculated cohesive energy  of carbon in graphene is much larger, 7.4 eV/C-atom \cite{win98} but that refers to typically carbon inside the skeleton and carbon at the edge will be less strongly bound. \cite{chu10} find a theoretical value of 5.4 eV for the 
C atoms at a zig-zag edge. Moreover, theoretical cohesive energies are not a good measure for the Arrhenius energy in unimolecular dissociation experiments \cite{bae96}. Given the low abundance of gas phase carbon, the reverse reaction is unimportant and the chemical lifetime ($\tau_{chem}^{-1}=k_{frag}(E)$) has to be compared to the dynamical lifetime of the region. We have evaluated this chemical lifetime for a typical internal energy of 10eV as a function of the size of the graphene sheet, $N_c$, and the results are shown in Fig.~4.

\subsection{Application to NGC 7023}

To evaluate graphene formation in NGC 7023 we look at PAH stability. The physical conditions in the nebula,
characterized by parameters $G_0$ and $n_H$, can be obtained (SI Text). As shown in Fig. 4, H loss is 
very efficient in NGC 7023 for PAHs up to 70 C atoms. We then evaluate the timescale for the loss of C 
atoms by the graphene flakes and compare this to the dynamical age of the nebula for a
$\simeq 70$ C atom PAH (Fig. 4). It appears that if we adopt $E_o=3.65$ eV, C-loss is rapid compared to 
the age of the nebula. 

\section{Discussion}

Our models reveal that 70 C atoms PAH become unstable to graphene formation in the studied region of NGC 7023. Much larger PAHs will survive closer to the star,
while smaller PAHs will rapidly be completely dehydrogenated further away but - because of size - this will not lead to the formation of fullerenes. 
As shown in laboratory experiments~\cite{chu10}, graphene sheets larger than about 70 C atoms can be transformed into fullerene, but in space this is driven by UV 
photons rather than energetic electrons. We surmise that  fullerene formation is initiated by single C atom loss~\cite{gir09} at the zig-zag 
edge of the graphene flake. Quantum chemical calculations indeed show that the armchair structure of graphene clusters 
stabilizes through in plane $\pi$-bond formation, while the open-shell orbitals associated with the dangling bonds are located at the zig-zag 
edge~\cite{win98}, making these atoms more labile~\cite{chu10}. A detailed study of graphene edge structure ~\cite{kos09}, based on 
experimental results~\cite{gir09}, has indeed shown that single C atom loss leads to the formation of reconstructed edges bearing pentagons. These defects dramatically 
stress graphene flakes, inducing significant positive curvature in its topology~\cite{she10}. Eventually, the barrier-less 
migration of the pentagons within the hexagonal network will lead to the zipping-up of the flake into C$_{60}$ (\cite{chu10}, Fig. 3).
The above mechanism is severely limited by the physical conditions, controlling dehydrogenation, and the size of the precursors, 
which need to be above $\sim$ 70 C atoms. As shown by our observations, only about 1$\%$ of the available PAHs are converted 
into C$_{60}$ by this process (Fig. 2). The activation of this ``top-down'' chemistry requires the high UV fields available near massive stars. As a corollary, high abundance of 
C$_{60}$ in the diffuse interstellar medium~\cite{foi94} reflects processing by massive stars, which are capable of dehydrogenating PAHs 
in the relevant size range (60-70 C atoms). PAHs much larger than than 70 C atoms will survive even in close encounters
with massive stars. Smaller size PAHs ( $<$ 50 C atoms) cannot reach the fullerene island of stability, hence their photo-processing (Fig. 2)
must be a source various carbon nano-compounds e.g. rings, chains, cages, bowls, tubes etc ( Fig. 3). Laboratory experiments have indeed shown
that processing of small graphene constrictions can lead to carbon rings and chains~\cite{chu09}. Finally, we note that recent studies suggest that PAH molecules
do not provide a satisfactory explanation to the diffuse interstellar bands \cite{sti11,xia11}. The compounds described in this top-down
chemistry may be relevant to understand these features of the UV extinction curve, in the Milky-Way and in other galaxies \cite{li08}.

\section{Conclusion}

Analyzing infrared observations of the NGC 7023 nebula we have found evidence that $C_{60}$ is being formed in the ISM. 
Classical bottom-up formation routes fail to explain these observations so we propose a new chemical route in which $C_{60}$ 
is formed directly by the photochemical processing of large PAH molecules. Other carbonaceous compounds (cages, tubes, bowls) 
are also the product of this photoprocessing. This route must be relevant in the interstellar medium, but may also be important in 
the inner regions of protoplanetary disks around solar-type stars where accretion on the young star generates intense UV fields and 
PAHs are known to be present. There, they can be converted, on timescale of $\sim$ 1 Myr, into fullerenes and hydrocarbon fragments. 
This source of organic compounds remains to be considered in models studying the organic photochemistry of the interstellar medium and 
regions of terrestrial planet formation.  It is clear that such studies may benefit greatly from the progress achieved  in the field 
of graphene stability.

\vspace{0.5cm}
{\small
{\bf Aknowledgements} Studies of interstellar PAHs at Leiden Observatory are supported through advanced-ERC grant 246976 from the 
European Research Council. The authors wish to thank the referees for their constructive comments. Laure Cadars
(www.laurecadars.com) is acknowledged for producing the C$_{60}$ formation sketch.}

%\end{article}
\begin{figure*}
\includegraphics[width=13cm]{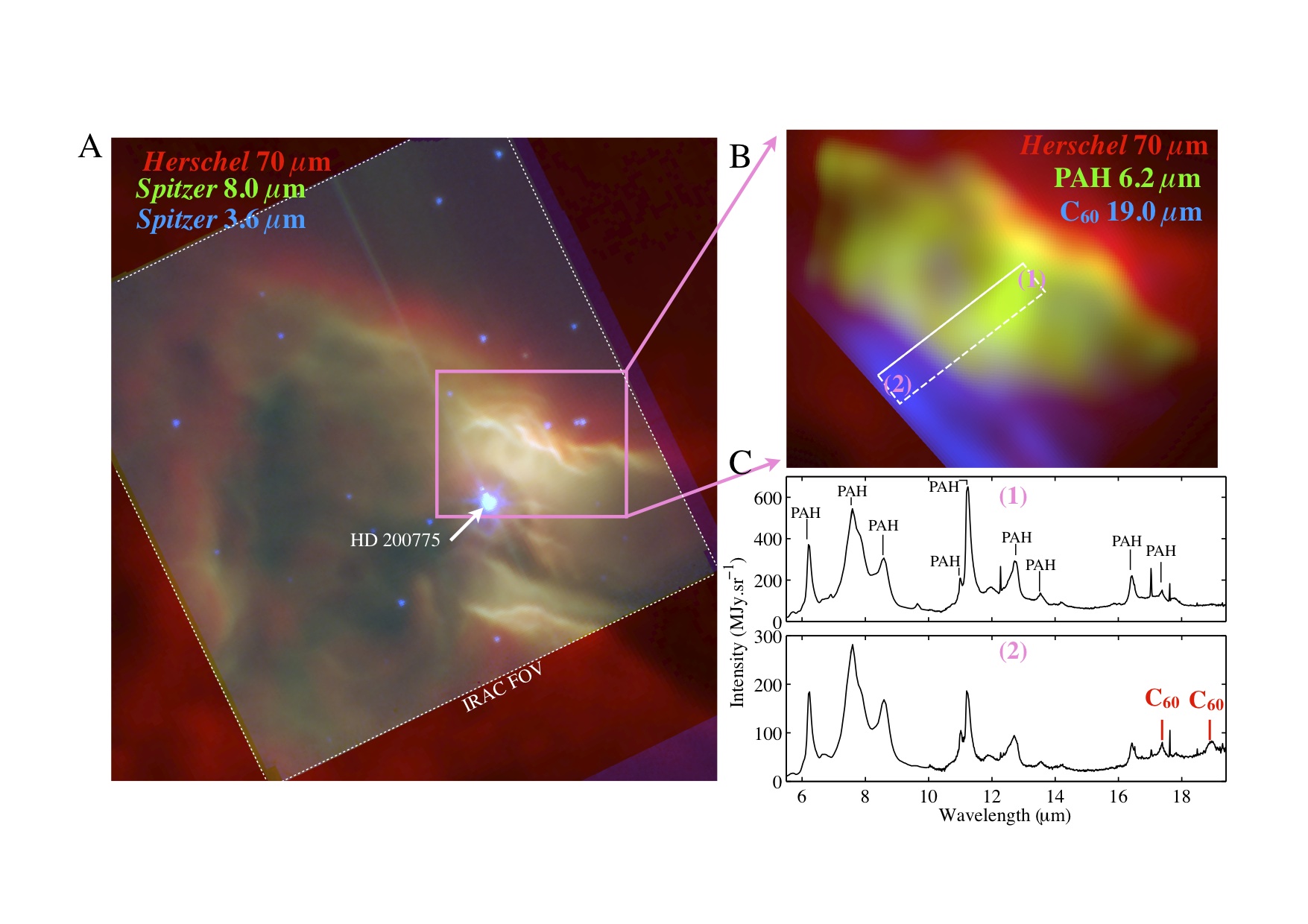}\\
\vspace{-1.5cm}
\caption{Overview of the NGC 7023 nebula. {\bf (A)} Multi wavelength color coded view of the nebula in the infrared. Red is the emission at 70 $\mu$m observed with the Photodetector Array Camera end Spectrometer (PACS) onboard \emph{Herschel}. This emission is dominated by dust. Green is the \emph{Spitzer}-Infrared Array Camera (IRAC)  8 $\mu$m emission tracing the PAH C-C mode and blue is the IRAC 3.6 $\mu$m emission, tracing the PAH C-H mode and stellar emission. The position of the intermediate mass young star HD 200775 illuminating the nebula is indicated. {\bf (B)}  Color coded image of the spatial distribution of different compounds in NGC 7023 : Red is the emission of dust observed with \emph{Herschel}-PACS, green shows the emission integrated in the 6.2 $\mu$m C-C band of PAHs observed with \emph{Spitzer}-Infrared Spectrograph (IRS), and blue is the C$_{60}$ emission observed with \emph{Spitzer}-IRS integrated in the 19.0 $\mu$m band. The white rectangle shows the region on which the extraction of the PAH and C$_{60}$ abundance (Figure 2) was performed. {\bf (C)} \emph{Spitzer} IRS mid-infrared spectra taken at positions (1) (upper panel) and (2) (lower panel) in {\bf (B)}. 
The distance from the star at positions (1) and (2) are respectively $\sim 35''$ and $\sim 15''$. The bands of PAHs and C$_{60}$ are labeled in the spectra.}
\end{figure*}

\begin{figure*}
\includegraphics[width=9cm, angle=0]{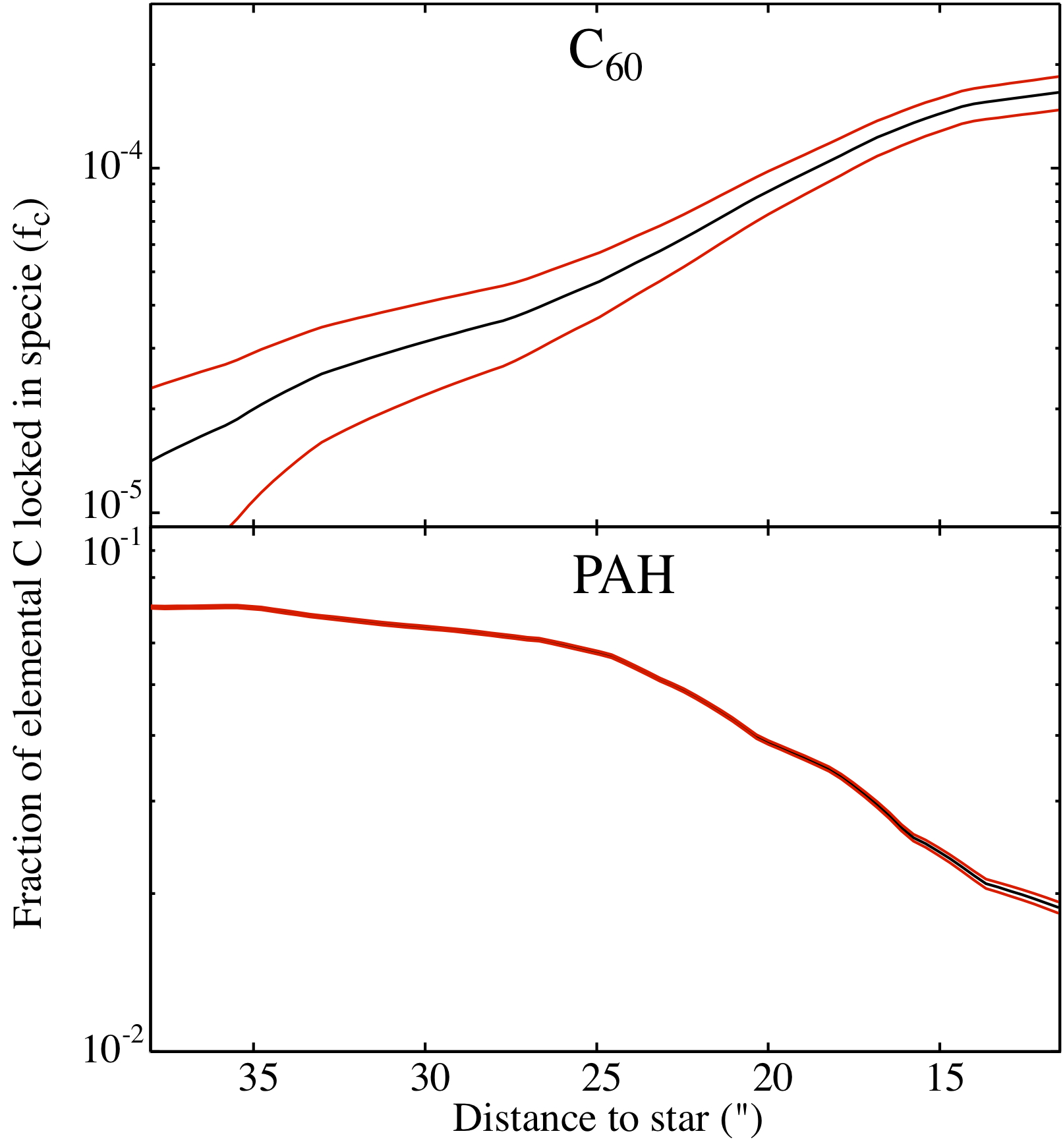}\\
\caption{Abundances of C$_{60}$ {\bf(A)} and  PAHs  {\bf(B)} in NGC 7023 as a function of distance from the star in the cut shown in Figure 1.
The  red curves give the 1 sigma uncertainty, obtained from the propagation of instrumental uncertainty on the extraction of integrated intensities of PAHs and C$_{60}$ bands.}
\end{figure*}

\begin{figure*}
\begin{center}
\includegraphics[width=11cm, angle=0]{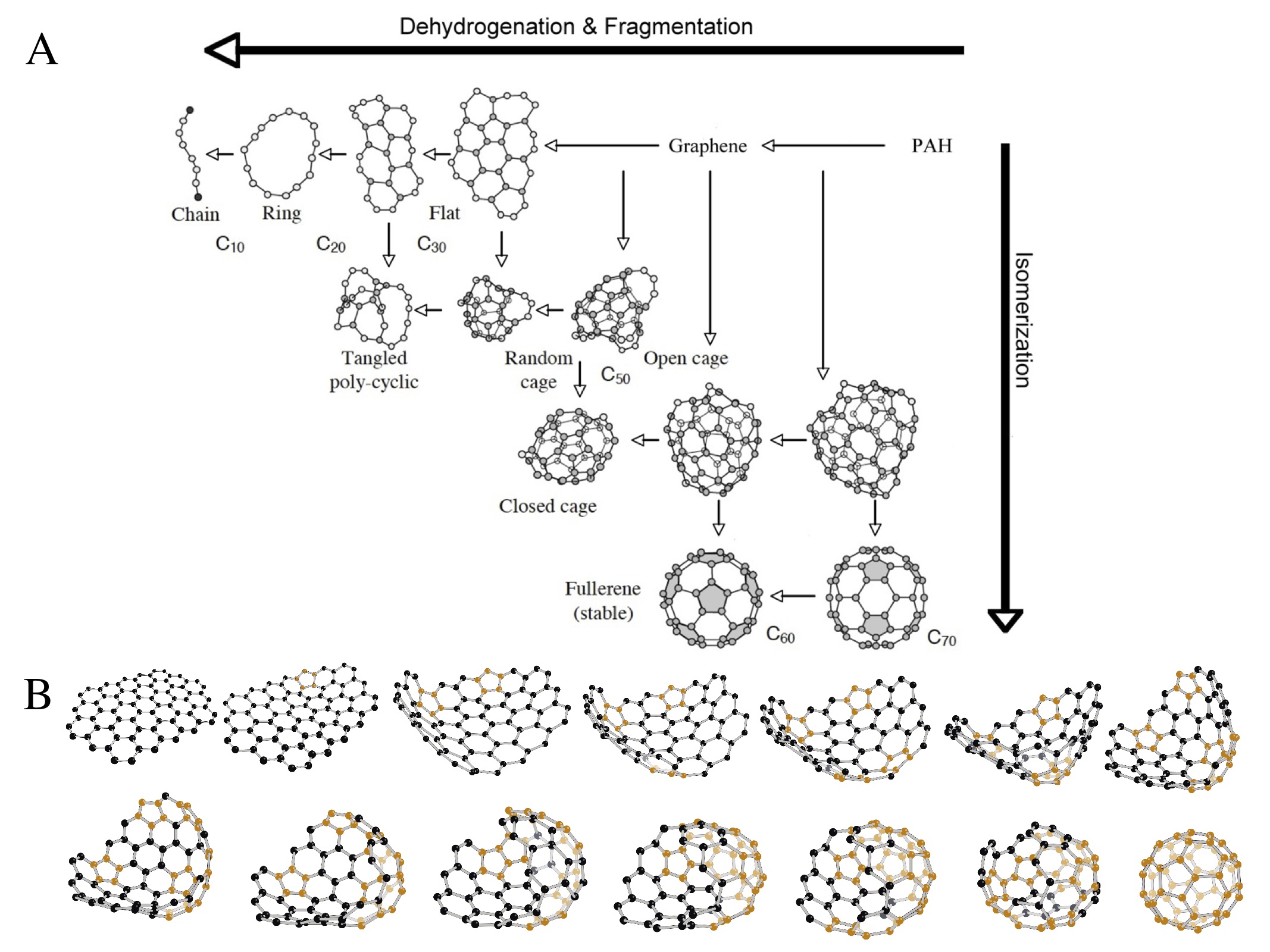}\\
\end{center}
\caption{Schematic representation of top-down interstellar carbon chemistry. {\bf (A)} The chemical evolution of PAHs in the interstellar medium under the influence of UV photons combines the effects of dehydrogenation and fragmentation with those of isomerization. Fully hydrogenated PAHs -- injected by stars into the ISM -- are at the top right side. Near bright stars, UV photolysis will preferentially lead to complete H-loss (e.g., the ``weakest link'') and the formation of graphene. Further fragmentation may lead to the formation of flats, rings, and chains. However, this process competes with Êisomerization to various types of stable intermediaries such as cages and fullerenes. {\bf (B)} Schematic illustration of conversion of graphene into C$_{60}$ in 13 steps.  Dehydrogenated PAHs, i.e. graphene sheets, loose carbon atoms under UV irradiation, giving rise to pentagonal defects (represented with orange C atoms) at the edges of the sheet. These defects in the hexagonal network induce curvature of the sheet, the migration of the pentagons allows the molecule to close. Image courtesy of L. Cadars (www.laurecadars.com).}
\end{figure*}

\begin{figure*}
\begin{center}
\includegraphics[width=12cm, angle=0]{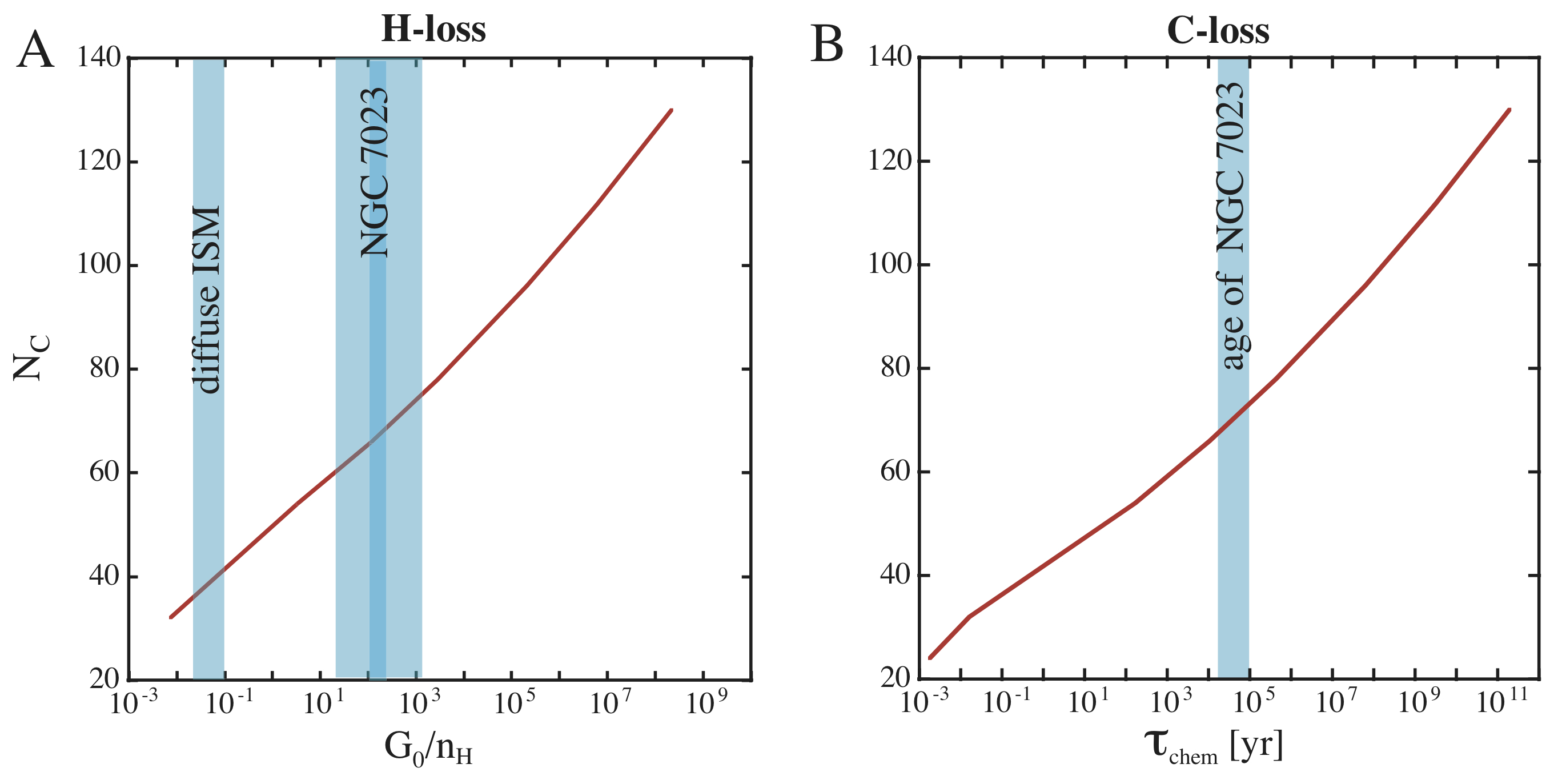}\\
\caption{Photochemistry of PAHs in NGC 7023 {\bf (A)} The red curve shows the evolution of the critical value when PAHs have lost half of their H atoms, as a function of the physical conditions ($G_0/n_H$ where $G_0$ is the radiation field in Habing units and $n_H$ is the density of H atoms in the gas in cm$^{-3}$, see Online Supporting Text for details) and the number of carbon atoms in the PAH molecule $N_C$. Above this line PAHs are stable against dehydrogenation. Below this curve, PAHs rapidly loose all their H atoms. The value of $G_0/n_H$ for NGC 7023 is shown, dark blue represents the ``narrow'' range and clear blue the ``broad'' range of values for this parameter (Online Supporting Text).   {\bf (B)} Time scale for the loss of a carbon atom by UV photon absorption as a function of graphene sheet size.}
\end{center}
\end{figure*}

\clearpage
\section*{Supporting Text}

\subsection*{Physical conditions in NGC 7023}

\subsubsection*{General properties}

NGC 7023 is a widely studied photodisociation region. We briefly review the information
on physical conditions available in the literature. The nebula has a hourglass-shaped, low-density 
cavity, which was opened in the dense molecular cloud by the winds of the young Be star (Fuente et al. 1998). 
The formation of C$_{60}$ is seen to occur inside this cavity (See Fig. 1 in the main text) in regions
particularly close to the star (10'' to 40"). In the next sections we discuss the values we adopt for 
radiation field and gas density in the cavity.

\subsubsection*{Radiation field}
The intensity of the radiation field in the cavity, $G_{0}$, can be derived based on the spectral  
type and flux of the star diluted by the square 
of the distance from the star. Doing this exercise, Joblin et al. (2010) find that the radiation 
field is $G_{0}=2600$ at a distance of 40'' from the star (similar to the value found by 
Rogers et al. 1995 and Chokshi et al. 1988). 
Using this value and spherical dilution we derive the range of radiation fields covered in the
cut where we study C$_{60}$ and PAH evolution (Figs. 1 and 2). We find $G_0=3.0\times10^4$ 
at the peak of $C_{60}$ abundance (12'' from the star) and $G_0=3.4\times10^3$ at the minimum
of C$_{60}$ abundance (35'' from the star). 
Fuente et al. (1999) have reported the highest value of radiation field in the PDR with $G_0=1\times10^{4}$ 
at the position of the $H_{2}$ filaments at 46'' from the star. This would correspond to values of
$G_0=1.7\times10^{4}$ and $G_0=1.5\times10^5$ at the edges of the cut (resp. 35'' and 12'' from 
the star). We therefore adopt average values $G_0=1.0\pm 0.7 \times 10^4$ and 
$G_0=1.0 \pm 0.7\times10^5$ at 35 and 12'' from the star from the star.

\subsubsection*{Density in the cavity}

The density of the atomic gas in this region is difficult to derive directly, but can be constrained 
from observations of the surrounding molecular cloud. More specifically, the density derived from 
CO lines by Gerin et al. (1998) in the back wall of the cavity, point to a value of $n_H^{Mol}\sim3000$ 
cm$^{-3}$. The atomic gas well within the cavity, is expected to be at least order of magnitude hotter, 
and hence an order of magnitude less dense if we consider pressure equilibrium, i.e. $n_H^{Cav}\sim 300$. 
Rogers et al. (1995) find that $n_{H}^{Mol}/n_{H}^{Cav}=10-35$, which then implies $n_{H}^{Cav}=85-300$. 
Joblin et al. (2010) quote a value of 100 cm$^{-3}$ for $n_{H}^{Cav}$. A more direct estimation of the 
column density of warm atomic gas can be derived from the dust emission. Since we have seen that 
dust temperature increases when getting closer to the star (see first section of this document), this
implies that this emission indeed comes form the cavity and not from the wall behind. Using the DUSTEM
(Compiegne et al. 2011) model, we can reproduce the observed emission in the PACS 70 $\mu$m 
filter at a distance of 35'' from the star ($\sim$ 1 Jy/pixel or 2300 MJy.sr$^{-1}$), for a radiation field 
of $G_0=10^{4}$, and leaving the column density as a free parameter. This yields $N_H=2\times10^{19}$ 
cm$^{-2}$ for the cavity, which with a physical size of $5.7\times10^{17}$cm (2$\times$35'' at 400 pc) 
corresponds to $n_{H}^{Cav}=45$cm$^{-2}$, somewhat lower than other estimates based on PDR 
or molecular tracers. Note that this determination does not depend significantly on the accuracy of 
determination of radiation field in the considered range of $G_0$ (see Compiegne et al 2011 Fig. 7).
We keep $n_{H}^{Cav}=50$ as lower limit and $n_{H}^{Cav}=250$ as upper limit and adopt 
$n_H^{Cav}=150\pm100$ cm$^{-3}$.

\subsubsection*{$G_{0}/n_{H}$}

Using the adopted average density and radiation field values we can obtain $G_{0}/n_{H}$ needed to 
estimate the dehydrogenation efficiency (see section ``Graphene formation'' in the manuscript). We find $G_{0}/n_{H}=65\pm45$ at 35'' 
from the star and $G_{0}/n_{H}=650\pm450$ at 12'' from the star. From these numbers we define the 
``broad'' and ``narrow'' domains of values of $G_{0}/n_{H}$ expected in the cut in NGC 7023, 
respectively [20-1100] and [105-200]. These are represented graphically in Fig. 4 over the dehydrogenation
stability curve.\\

\clearpage
\subsection*{Supplementary videos}
\vspace{1cm}
\begin{center}{https://sites.google.com/site/olivierberne/c60-formation-from-graphene} \end{center}
\noindent {S-Video 1:} Conversion of graphene into C$_{60}$. This video shows schematically  how graphene can be converted into C$_{60}$. We start with a sheet of graphene where carbon atoms are arranged in a hexagonal network. Under UV irradiation, C atoms are lost in the hexagons situated at the edge of the graphene flake and converted into pentagons (shown in orange). The formation of pentagons stresses the molecule forcing it to curve. When several pentagons have been formed they can migrate inside the molecule. Finally, when 12 pentagons have been formed and when each carbon atom belongs to both a pentagon and a hexagon, the molecule closes into C$_{60}$.

\clearpage
\subsection*{Supplementary figures}

\begin{figure*}[!h]
\includegraphics[width=10cm]{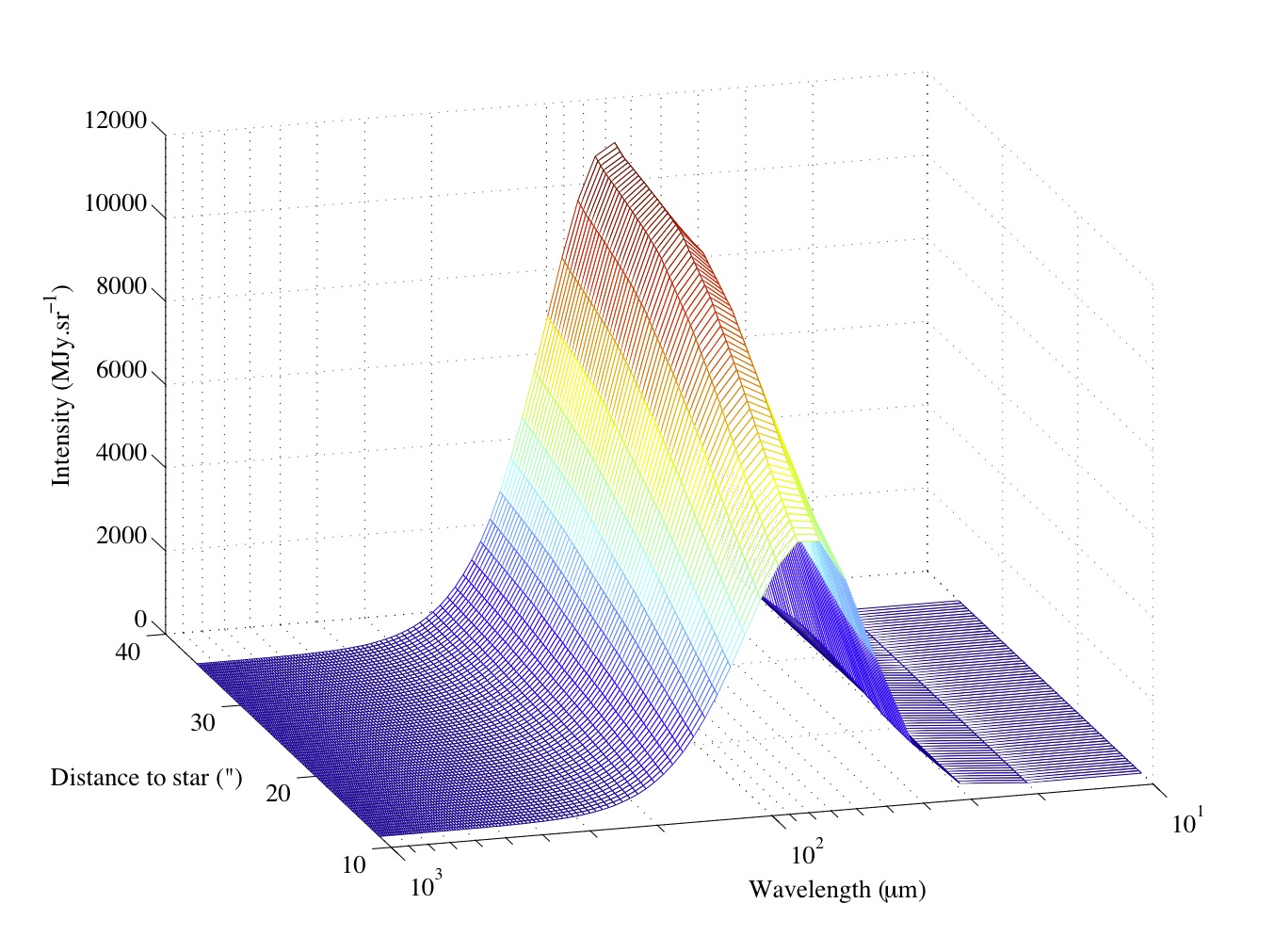}
{\\ Figure S-1: Fitted modified black bodies to the SEDs observed with Herschel in the cross cut from position (1) to (2) show in Fig. 1.}
\end{figure*}

\begin{figure*}[!h]
\includegraphics[width=13cm]{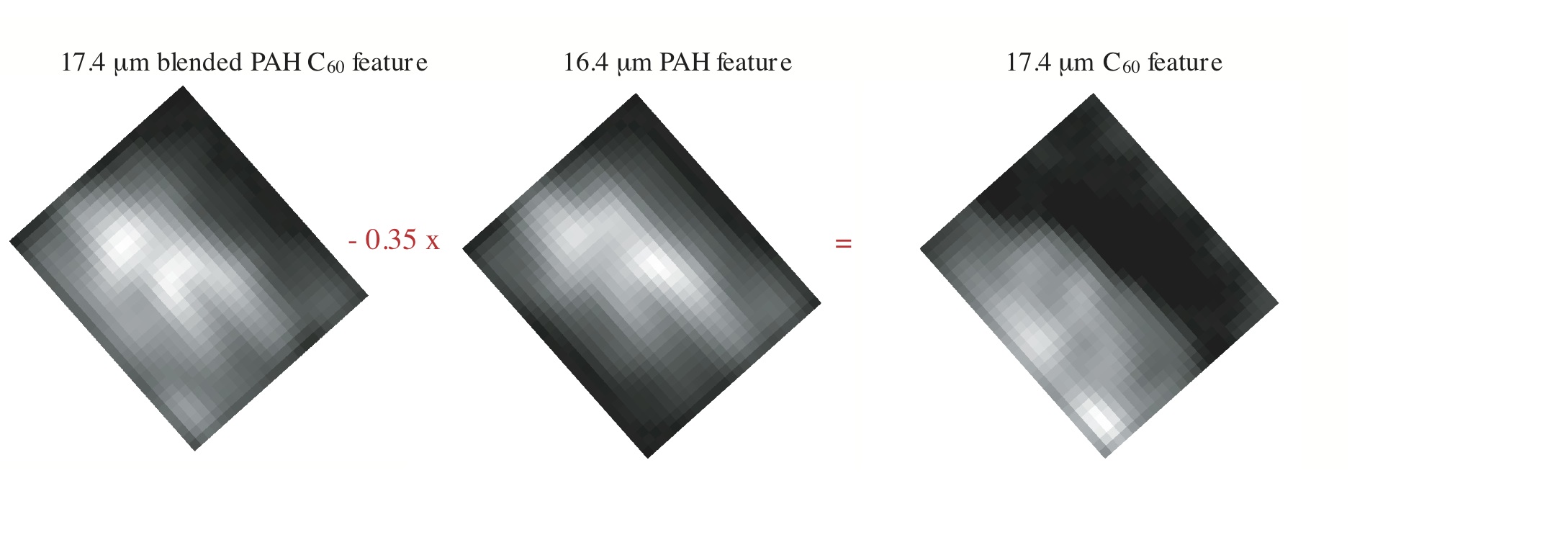}
{\\ Figure S-2: Subtraction of the contamination of PAHs in the 17.4 $\mu$m band.}
\end{figure*}

\begin{figure*}[!h]
{\includegraphics[width=0.7\textwidth]{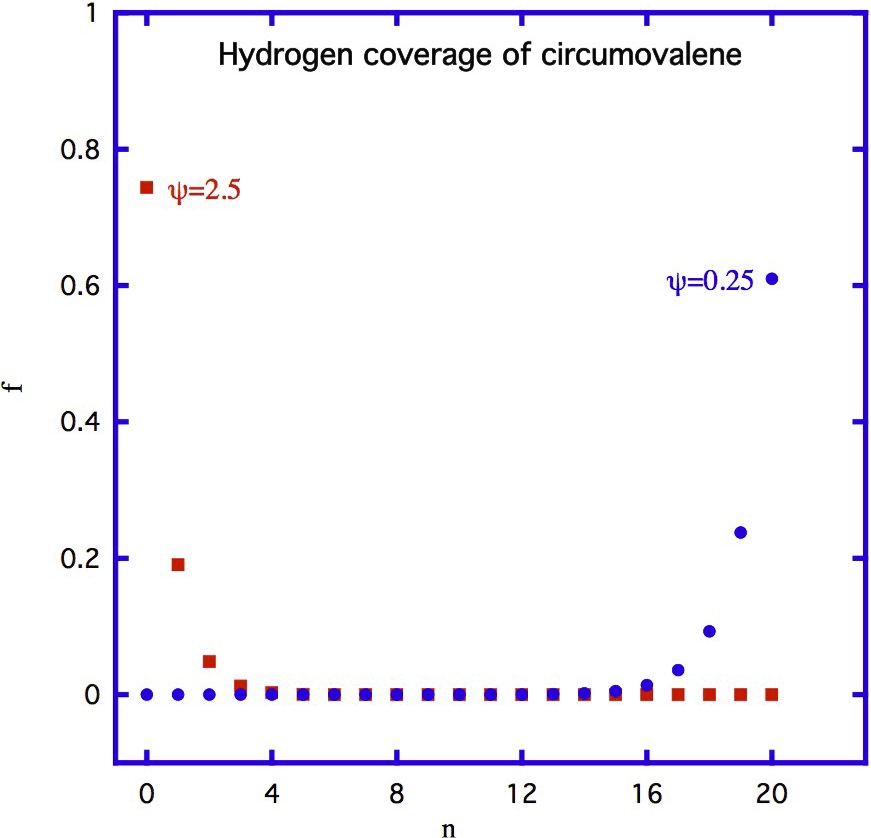}} 
{\\ Figure S-3 The relative abundance of circumovalene and its partial dehydrogenated derivatives, C$_{66}$H$_{n}$, for two values of the dissociation parameter, $\psi$.} 
\label{fig:PAHs2graphene}
\end{figure*}

{}


\begin{thebibliography}{}

\bibitem{tie08}
{{Tielens}, A.~G.~G.~M.}
{\em {Interstellar Polycyclic Aromatic Hydrocarbon
  Molecules}}.
 \emph{{Ann. Rev. Astron. Astrophys.}}
  \textbf{{46}}, pp.~{289--337}
  ({2008}).





\bibitem{cam10}
{{Cami}, J.}, {{Bernard-Salas}, J.},
  {{Peeters}, E.} \& {{Malek}, S.~E.}
{\em {Detection of C$_{60}$ and C$_{70}$ in a Young
  Planetary Nebula}}.
 \emph{{Science}} \textbf{{329}},
pp.~{1180--1182} ({2010}).

\bibitem{sel10}
{{Sellgren}, K.} \emph{et~al.}
{\em {C$_{60}$ in Reflection Nebulae}}.
 \emph{{Astrophys. J.}}
  \textbf{{722}}, pp.~{54--57}
  ({2010}).
  
  \bibitem{kro85}
{{Kroto}, H.~W.}, {{Heath}, J.~R.},
  {{Obrien}, S.~C.}, {{Curl}, R.~F.} \&
  {{Smalley}, R.~E.}
{\em {C(60): Buckminsterfullerene}}.
 \emph{{Nature}} \textbf{{318}}, ({1985})
pp.~{162--163} .

\bibitem{goe92}
{{Goeres}, A.} \& {{Sedlmayr}, E.}
{\em {The envelopes of R Coronae Borealis stars. I - A
  physical model of the decline events due to dust formation}}.
 \emph{{Astron. Astrophys.}}
  \textbf{{265}}, pp.~{216--236}
  ({1992}).

\bibitem{che00}
{{Cherchneff}, I.}, {{Le Teuff}, Y.~H.},
  {{Williams}, P.~M.} \& {{Tielens},
  A.~G.~G.~M.}
{\em {Dust formation in carbon-rich Wolf-Rayet stars. I.
  Chemistry of small carbon clusters and silicon species}}.
 \emph{{Astrophys. J.}}
  \textbf{{357}}, pp.~{572--580}
  ({2000}).

\bibitem{pas00}
{{Pascoli}, G.} \& {{Polleux}, A.}
{\em {Condensation and growth of hydrogenated carbon
  clusters in carbon-rich stars}}.
 \emph{{Astron. Astrophys.}}
  \textbf{{359}}, pp.~{799--810}
  ({2000}).

\bibitem{kro88}
{{Kroto}, H.~W.} \& {{McKay}, K.}
{\em {The formation of quasi-icosahedral spiral shell
  carbon particles}}.
 \emph{{Nature}} \textbf{{331}},
pp.~{328--331} ({1988}).

\bibitem{hea92}
{{Heath}, J.~R.}
{\em {Synthesis of C$_{60}$ from Small Carbon
  Clusters}}, chap.~{2}, pp.~{1--23}
  ({1992}).

\bibitem{hun98}
{Hunter, J.~M.}, {Fye, J.~L.},
  {Roskamp, E.~J.} \& {Jarrold, M.~F.}
{\em Annealing carbon cluster ions: A mechanism for
  fullerene synthesis}.
 \emph{{J. Phys. Chem.}}
  \textbf{{98}}, pp.~{1810--1818}
  ({1994}).

\bibitem{irl06}
{Irle, S.}, {Zheng, G.}, {Wang,
  Z.} \& {Morokuma, K.}
{\em The C$_{60}$ formation puzzle â ``solved'' qm/md
  simulations reveal the shrinking hot giant road of the dynamic fullerene
  self-assembly mechanism}.
 \emph{{J. Phys. Chem. B}}
  \textbf{{110}}, pp.~{14531--14545}
  ({2006}).

\bibitem{wer04}
{{Werner}, M.~W.} \emph{et~al.}
{\em {The Spitzer Space Telescope Mission}}.
 \emph{{Astrophys. J.}}
  \textbf{{154}}, pp.~{1--9} ({2004}).

\bibitem{pil10}
{\em {Pilbratt}, G.~L.} \emph{et~al.}
{{Herschel Space Observatory. An ESA facility for
  far-infrared and submillimetre astronomy}}.
 \emph{{Astron. Astrophys.}}
  \textbf{{518}}, ({2010}).

\bibitem{rap05}
{{Rapacioli}, M.}, {{Joblin}, C.} \&
  {{Boissel}, P.}
{\em {Spectroscopy of polycyclic aromatic hydrocarbons and
  very small grains in photodissociation regions}}.
 \emph{{Astron. Astrophys.}}
  \textbf{{429}}, pp.~{193--204}
  ({2005}).

\bibitem{ber07}
{{Bern{\'e}}, O.} \emph{et~al.}
{\em {Analysis of the emission of very small dust
  particles from Spitzer spectro-imagery data using blind signal separation
  methods}}.
 \emph{{Astron. Astrophys.}}
  \textbf{{469}}, pp.~{575--586}
  ({2007}).

\bibitem{bre05}
{{Br{\'e}chignac}, P.} \emph{et~al.}
{\em {Photoinduced products from cold coronene clusters. A
  route to hydrocarbonated nanograins?}}
 \emph{{Astron. Astrophys.}}
  \textbf{{442}}, pp.~{239--247}
  ({2005}).

\bibitem{ros11}
{Rosenberg, M.}, {Bern\'e, O.},
  {Boersma, C.}, {Allamandola, L.} \&
  {Tielens, A. G. G.~M.}
{\em {Coupled Blind Signal Separation and Spectroscopic
  Database Fitting of the Mid Infrared PAH Features }}.
 \emph{{Astron. Astrophys.}} , {532, pp.~128--142}, (2011)




\bibitem{job03}
{{Joblin}, C.}
{\em {Carbon macromolecules in the cycle of interstellar
  matter: observational and laboratory experiments}}.
 In{{F.~Combes, D.~Barret, T.~Contini, \& L.~Pagani}}
  (ed.) \emph{{SF2A-2003: Semaine de l'Astrophysique
  Francaise}}, pp.~{175--179} ({2003}).

\bibitem{chu10}
{{Chuvilin}, A.}, {{Kaiser}, U.},
  {{Bichoutskaia}, E.}, {{Besley}, N.~A.} \&
  {{Khlobystov}, A.~N.}
{\em {Direct transformation of graphene to fullerene}}.
 \emph{{Nature Chemistry}}
  \textbf{{2}}, pp.~{450--453}
  ({2010}).

\bibitem{pet05}
{{Pety}, J.} \emph{et~al.}
{\em {Are PAHs precursors of small hydrocarbons in
  photo-dissociation regions? The Horsehead case}}.
 \emph{{Astron. \& Astrophys.}}
  \textbf{{435}}, pp.~{885--899}
  ({2005}).





\bibitem{pog10}
Poglitsch, A.; Waelkens, C.; Geis,
{\em The Photodetector Array Camera and Spectrometer (PACS) on the Herschel Space Observatory}
\emph{Astron. Astrophys.},  {518},  (2010)

\bibitem{gri10}
Griffin, M. J.; Abergel, A.; Abreu et al.
{\em The Herschel-SPIRE instrument and its in-flight performance}
\emph{Astron. Astrophys.},  {518},  (2010)\


\bibitem{abe10}
Abergel, A.; Arab, H.; Compigne, M., 
{\em Evolution of interstellar dust with Herschel. First results in the photodissociation regions of NGC 7023}
\emph{Astron. Astrophys.},  {\bf 518}, pp.~96--100 (2010)

\bibitem{ber08}
Bern\'e, O. 
{\em Evolution des tr\`es petites particules de poussi\`ere dans le cycle cosmique de la mati\`ere : m\'ethodes de s\'eparation aveugle de sources et spectro-imagerie avec le t\'elescope spatial Spitzer.}
PhD Thesis, Universit\'e Paul Sabatier, http://thesesups.ups-tlse.fr/381/ pp.~148-153  (2008) 

\bibitem{tie05}
{{Tielens}, A.~G.~G.~M.}
{\em {{The Physics and Chemistry of the Interstellar
  Medium}}} ({2005}).


\bibitem{job10}
{{Joblin}, C.} \emph{et~al.}
{\em {Gas morphology and energetics at the surface of
  PDRs: New insights with Herschel observations of NGC 7023}}.
 \emph{{Astron. Astrophys.}}
  \textbf{{521}}, pp.~{25--29} ({2010}).

\bibitem{eke97}
{{Ekern}, S.~P.}, {{Marshall}, A.~G.},
  {{Szczepanski}, J.} \& {{Vala}, M.}
{\em {Photon-induced Complete Dehydrogenation of Putative
  Interstellar Polycyclic Aromatic Hydrocarbon Cations: Coronene and
  Naphtho[2,3-a]pyrene}}.
 \emph{{Astrophys. J.}}
  \textbf{{488}}, pp.~{39--42} ({1997}).

\bibitem{joc94}
{{Jochims}, H.~W.}, {{Ruhl}, E.},
  {{Baumgartel}, H.}, {{Tobita}, S.} \&
  {{Leach}, S.}
{\em {Size effects on dissociation rates of polycyclic
  aromatic hydrocarbon cations: Laboratory studies and astophysical
  implications}}.
 \emph{{Astrophys. J.}}
  \textbf{{420}}, pp.~{307--317}
  ({1994}).

\bibitem{nov05}
Novoselov et al. {\em Two-dimensional atomic crystals} Proc. Nat. Ac. Sci. 102, pp.~10451--10453




\bibitem{lep03}
{{Le Page}, V.}, {{Snow}, T.~P.} \&
  {{Bierbaum}, V.~M.}
{\em {Hydrogenation and Charge States of Polycyclic
  Aromatic Hydrocarbons in Diffuse Clouds. II. Results}}.
 \emph{{Astrophys. J.}}
  \textbf{{584}}, pp.~{316--330}
  ({2003}).

\bibitem{mon11}
{{Montillaud}, J.}, {{Joblin}, C.} \&
  {{Toublanc}, D.}
{\em {Modelling the physical and chemical evolution of
  PAHs and PAH-related species in astrophysical environments}}.
 In \emph{{EAS Publications Series}},
  vol.~{46} of \emph{{EAS Publications
  Series}}, pp.~{447--452} ({2011}).
  
\bibitem{she10}
V.~B. Shenoy, C.~D. Reddy, Y.-W. Zhang, 
{\em Spontaneous Curling of Graphene Sheets with Reconstructed Edges} { ACS Nano} {\bf 4}, pp.~4840--4844 (2010).

\bibitem{klo89}
Klots, C. E.
{\em Thermal kinetics in small systems}
\emph{J. Chem. Phys.} {\bf 90}, pp.~4470-4472

\bibitem{bae96}
Baer, T. and Hase W. L.  {\em Unimolecular reaction dynamics : Theory and experiment}, Oxford : Oxford University Press (1996)

\bibitem{hab68}
{{Habing}, H.~J.}
{\em {The interstellar radiation density between 912 A and
  2400 A}}.
 \emph{{Bull. Astron. Inst. Neth.}}
  \textbf{{19}}, pp.~{421}
  ({1968}).

\bibitem{sno98}
Snow, Theodore P.; Le Page, Valery; Keheyan, Yeghis; Bierbaum, Veronica M.
{\em The interstellar chemistry of PAH cations}
\emph{Nature} {\bf 391} pp.~259--261 (1998)

\bibitem{mic10}
Micelotta, E. R.; Jones, A. P.; Tielens, A. G. G. M.
{\em Polycyclic aromatic hydrocarbon processing in a hot gas}
\emph{Astron. Astrophys.} {\bf 510}, A37 (2010)


\bibitem{win98}
{Winter, N.} \& {Ree, F.}
{\em Carbon particle phase stability as a function of
  size}.
 \emph{{J. Comp. Mat. Des.}}
  \textbf{{5}}, pp.~{279--294}
  ({1998}).

\bibitem{kos09}
{Koskinen, P.}, {Malola, S.} \&
  {H\"akkinen, H.}
{\em Evidence for graphene edges beyond zigzag and
  armchair}.
 \emph{{Phys. Rev. B}} \textbf{{80}},
pp.~{73401--73404} ({2009}).

 
 
\bibitem{gir09}
Girit, c {\it et~al.\/}, {\em Graphene at the edge} { Science\/} {\bf 323}, pp.~1705--1708 (2009). 


  
  \bibitem{foi94}
{{Foing}, B.~H.} \& {{Ehrenfreund}, P.}
{\em {Detection of two interstellar absorption bands
  coincident with spectral features of C$_{60}^{+}$}}.
 \emph{{Nature}} \textbf{{369}},  
pp.~{296--298} ({1994}).

\bibitem{chu09}
A.~Chuvilin, J.~C. Meyer, G.~Algara-Siller, U.~Kaiser, {\em From Graphene constrictions to single carbon chains} {New Journal of
  Physics} {\bf 11}, 83019 (2009).

\bibitem{xia11}
Xiang, F. Y.  Li, A.; Zhong, J. X. {\em A Tale of Two Mysteries in Interstellar Astrophysics: The 2175 $\AA$ Extinction Bump and Diffuse Interstellar Bands }
Astrphys. J, {\bf 733} pp 91-100 (2011)

\bibitem{sti11}
Steglich, M. Bouwman, J. Huisken, F. Henning, T. {\em Can neutral and ionized PAHs be carriers of the UV extinction bump and the diffuse interstellar bands?}
eprint arXiv:1108.2972 (2011)

\bibitem{li08}
Li, A. Chen, J. H. Li, M. P. Shi, Q. J. Wang, Y. J. {\em On buckyonions as an interstellar grain component}
Mon. Not. R. Astron. Soc. {\bf 390} pp 39-42 (2008)

\end{thebibliography}

\clearpage


\begin{thebibliography}{}

%% Created for Olivier Berne at 2008-05-19 16:36:54 +0200 

%% Saved with string encoding Unicode (UTF-8) 


\bibitem[Chokshi et al.(1988)]{cho88}
Chokshi, A.; Tielens, A. G. G. M.; Werner, M. W.; Castelaz, M. W.
C II 158 micron and O I 63 micron observations of NGC 7023 - A model for its photodissociation region
\emph{Astrophys. J,} {\bf334} 803 (1988)

\bibitem[Compi\`egne et al.(2011)]{com11}
Compi\`egne, M.; Verstraete, L.; Jones, A.; Bernard, J.-P.; Boulanger, F.; Flagey, N.; Le Bourlot, J.; Paradis, D.; Ysard, N.
The global dust SED: tracing the nature and evolution of dust with DustEM
\emph{Astron. Astrophys.}, {\bf 525}, 103, (2011)

\bibitem[Fuente et al.(1998)]{fue98}
Fuente, A.; Martin-Pintado, J.; Bachiller, R.; Neri, R.; Palla, F.
Progressive dispersal of the dense gas in the environment of early-type and late-type Herbig Ae-Be stars
\emph{Astron. Astrophys.}, {\bf334}, 253 (1998) 


\bibitem[Fuente et al.(1999)]{fue99}
Fuente, A.; Mart'n-Pintado, J.; Rodr'guez-Fern‡ndez, N. J.; Rodr'guez-Franco, A.; de Vicente, P.; Kunze, D.
Infrared Space Observatory Observations toward the Reflection Nebula NGC 7023: A Nonequilibrium Ortho-to-Para-H$_2$ Ratio
\emph{Astrophys. J.}, {\bf 518}, 45 (1999)

\bibitem[Gerin et al.(1998)]{ger98}
Gerin, M.; Phillips, T. G.; Keene, J.; Betz, A. L.; Boreiko, R. T.
CO, C i, and C II Observations of NGC 7023
Astrophys. J. {\bf500}, 329, (1998)


\bibitem[Poglitsch et al.(2010)]{pog10}
Poglitsch, A.; Waelkens, C.; Geis,
The Photodetector Array Camera and Spectrometer (PACS) on the Herschel Space Observatory
\emph{Astron. Astrophys.},  {\bf 518}, 2 (2010)

\bibitem[Rogers et al.(1995)]{rog95}
Rogers, C.; Heyer, Mark H.; Dewdney, P. E.
H I, CO, and IRAS observations of NGC 7023
\emph{Astrophys. J.} {\bf 442}, 694 (1995)


\bibitem[Tielens(2005)]{tie05}
Tielens,ÊA.ÊG.ÊG.ÊM. The Physics and Chemistry of the Interstellar Medium, 
Cambridge, UK: Cambridge, (2005)


\end{thebibliography}
\end{document}